\documentclass[fleqn,usenatbib]{mnras}

\usepackage{newtxtext,newtxmath}

\usepackage[T1]{fontenc}
\usepackage{ae,aecompl}

\usepackage{graphicx}	
\usepackage{amsmath,amstext}	
\usepackage{amssymb}	
\usepackage{xspace}

\newcommand{\hst}{{\it HST}\xspace}
\newcommand{\galex}{{\it GALEX}\xspace}

\title[Hot populations in M31's globular clusters]{Hubble Space Telescope FUV observations of M31's globular clusters suggest a spatially homogeneous helium-enriched sub population}

\author[M. B. Peacock et al.]{
Mark B. Peacock,$^{1}$\thanks{E-mail: mpeacock@msu.edu (MBP)}
Stephen E. Zepf,$^{1}$
Thomas J. Maccarone$^{2}$
Arunav Kundu$^{3}$
\newauthor
Christian Knigge$^{4}$
Andrea Dieball$^{5}$
and Jay Strader$^{1}$
\\
$^{1}$Michigan State University, East Lansing, MI 48824, USA\\
$^{2}$Texas Tech University, Lubbock, TX 79409, USA\\
$^{3}$Eureka Scientific Inc., 2452 Delmer Street, Suite 100 Oakland, CA 94602, USA\\
$^{4}$University of Southampton, Southampton SO17 1BJ, UK\\
$^{5}$Universit{\"a}t Bonn, Auf dem H{\"u}gel 71, D-53121 Bonn, Germany
}

\date{Accepted XXX. Received YYY; in original form ZZZ}

\pubyear{2018}

\begin{document}
\label{firstpage}
\pagerange{\pageref{firstpage}--\pageref{lastpage}}
\maketitle

\begin{abstract}
We present high spatial resolution, far ultraviolet (FUV) F140LP observations of 12 massive globular clusters in M~31 obtained using the ACS/SBC on the Hubble Space Telescope (\hst). These observations resolve the cluster profiles to scales similar to their core radii and enable the study of the spatial distribution of blue and extreme horizontal branch (HB) stars, which dominate the emission in the F140LP images. We confirm that some of these clusters have excess FUV emission, suggesting additional hot populations beyond those expected by canonical single stellar populations models. We find no evidence that the hot populations are spatially distinct from the majority populations in these clusters, as would be expected if the excess FUV emission is a result of a dynamically enhanced population of extreme-HB stars. We conclude that a second population of stars with significantly enhanced helium abundance is a viable explanation for the observed FUV emission which is both bright and distributed similarly to the rest of the cluster light. Our results support the use of FUV observations as a path to characterising helium enhanced sub populations in extragalactic clusters. These M31 clusters also show a correlation such that more massive and denser clusters are relatively FUV bright. Similar to extant Milky Way results, this may indicate the degree of helium enrichment, or second population fraction increases with cluster mass.
\end{abstract}




\section{Introduction}
\label{sec:intro}

Far-ultraviolet (FUV) observations are a unique probe of hot stellar populations. In old systems, such as globular clusters and early-type galaxies, their main sequence and red giant branch stars are generally too cool to emit significantly at such wavelengths. Instead, their FUV emission is dominated by their much rarer hot populations.

A surprising result from early observations of passive, red early-type galaxies is that they are brighter in the FUV than expected; exhibiting a `UV-upturn' \citep[e.g.][]{Code69, Dorman95, OConnell99}. Recent studies, based on Galaxy Evolution Explorer (\galex) observations, have noted that only a fraction of early-type galaxies exhibit a true UV upturn, defined as increasing  $F_{\lambda}$ with $\lambda$ below $\sim$2000\AA\ \citep[][]{Yi11}. However, the vast majority still exhibit significant excess FUV emission compared to that expected based on their spectral energy distributions from the near-UV through infrared \citep{Smith12b}. The leading explanation for this excess FUV emission from old populations is the presence of extreme horizontal branch (HB) stars \citep{Greggio90, OConnell99, Brown00}. This hypothesis is supported by \hst observations of M~32, which favour a blue/extreme-HB star explanation for its UV upturn \citep{Brown08}. While HB stars are an expected phase of stellar evolution, the magnitude of the FUV excess and the origin of the hottest extreme-HB stars in metal rich environments is quite poorly understood. Several mechanisms have been proposed to produce these stars, including: high mass loss on the red giant branch at high metallicity \citep{Greggio90}; binary interactions \citep{Han07}; and helium enhanced sub populations \citep{Lee05, Chung11}.

Globular clusters are prime environments to study HB stars due to their similar ages and metallicities. Dynamical interactions between stars also effect the binary populations in the clusters, allowing us to study the influence of such interactions on the evolution of HB stars. Studies of the Galactic globular clusters have shown that they host complex HB star morphologies. The HB stars in lower metallicity clusters are, on average, bluer than those in metal rich clusters and metallicity has long been proposed as the `first parameter' effecting the HB stars \citep{Sandage60}. However, metallicity alone cannot explain the observations. For example, different HB star populations are observed in similar metallicity clusters \citep[e.g.][]{Bellazzini01}. Metallicity also appears to have less of an influence on the `tail' of the HB star morphology \citep[i.e. the presence of very hot, extreme-HB stars, e.g.][]{OConnell99} than it does on the fraction of stars redder or bluer than the instability strip. These complexities of the HB phase of stellar evolution led to the infamous `second parameter' problem and several additional parameters have been proposed to influence these stars, such as: age; helium abundance (and second population stars); stellar core rotation; and the stellar density in the cluster core \citep[for a discussion see e.g.][]{Catelan09}.

In recent years, increased helium abundance has been proposed as a natural explanation for the formation of bluer HB stars. This is due to the compelling evidence that most of the Galactic globular clusters are not true simple stellar populations, but rather host at least two populations, with the `second population' having significantly different abundances of light and intermediate mass elements \citep[see e.g. the reviews of][and references therein]{Gratton12, Bastian17}. Current theories for the formation of these multiple populations struggle to explain all of the observations \citep[e.g.][]{Bastian15}. However, most require that the second population stars are significantly enriched in helium \citep[e.g.][]{DAntona07, Bragaglia10}. Such enhanced helium fractions have been observed in spectroscopic observations of a small number of cluster stars \citep[e.g.][]{Pasquini11, Marino11, Marino14} and can explain the observed \hst colour magnitude diagrams of many clusters \citep[e.g.][]{Milone12b, Piotto13, Milone15b}. Based on current samples, the fraction of second populations stars and their helium enrichment appears to vary significantly in different clusters (with $0.01<\Delta Y<0.12$) and may correlate with cluster mass \citep[][]{WagnerKaiser16, Milone17}. The combination of first population stars and helium enhanced second population stars can help to explain the complex HB star morphologies observed in the Galactic globular clusters \citep[e.g.][]{Lee05, DAntona05, Joo13}.

Another potentially important mechanism for the formation of extreme-HB stars in the dense cores of globular clusters is dynamical formation. Several dynamical processes may produce these very hot HB stars. Firstly, close encounters may produce bluer HB stars both through enhanced mass loss from tidal stripping and via helium dredge up into the stellar atmosphere \citep[e.g.][]{Suda07, Pasquato14}. It is also thought that mass loss in binary interactions is an important mechanism for forming extreme-HB stars in the Galactic field \citep[aka subdwarf~B stars;][]{Maxted01, Han03}. The dense stellar environments in the cores of globular clusters are known to significantly influence binary properties \citep[e.g.][]{Heggie75} and may therefore influence the formation of extreme-HB stars. Additionally, the tightening of binaries via stellar interactions may cause a binary of two helium white dwarfs to merge -- producing an extreme-HB like star \citep{Han03, Han08}. If any of these processes are important in the production of extreme-HB stars, then the formation of these stars should be enhanced in the cores of clusters with higher stellar densities. Studies of the Galactic globular clusters have tried to test for this correlation, but claims there is a correlation \citep{Fusi_Pecci93, Buonanno97, Rich97, Dotter10} have been intermixed with those that suggest there may not be \citep[e.g.][]{VandenBergh93, Recio_blanco06, Dalessandro12}.

Recent modelling of HB stars has suggested that both of these effects -- enhanced mass loss due to dynamical interactions and helium enhancement -- may be required to produce the hottest extreme-HB stars \citep{Yaron17}.

Extending the study of HB stars to globular clusters beyond the Milky Way and its satellites enables us to improve our understanding of them by expanding the sample -- especially important for those that are rare in the Milky Way, such as massive and metal-rich clusters. To this end, we present in this paper new, high spatial resolution FUV observations of M31's globular clusters. At these wavelengths, a population of blue/extreme-HB stars will dominate the emission from globular clusters \citep[e.g.][]{Moehler01} -- this is because other hot populations are either orders of magnitude fainter than these stars (such as white dwarfs and cataclysmic variables) or they are too rare to dominate over such a population (such as post-AGB stars and low mass X-ray binaries). Observations of the Galactic globular clusters confirm that their integrated FUV colours correlate with their HB star morphology \citep[][]{Catelan09, Dalessandro12}.

In this paper we consider a sample of 12 massive and FUV bright globular clusters in M~31 (see Section \ref{sec:sample} for details), for which we present new high resolution \hst ACS/SBC F140LP observations (Section \ref{sec:data}). We derive total FUV magnitudes of these clusters and compare to published photometry based on lower signal to noise and lower spatial resolution \galex observations (Section \ref{sec:mag}). We also present the ultraviolet surface brightness profiles at both F140LP (FUV) and F336W ($u$-band) wavelengths (Section \ref{sec:sbp}), which we use to consider the distribution of hot HB stars in these clusters. We conclude by discussing the constraints these data place on dynamically enhanced extreme-HB star populations (Section \ref{sec:ehb_dyn}) and the evidence for second populations of helium enhanced stars in these clusters (Section \ref{sec:ehb_msp}).

\begin{table*}
	\centering
	\caption{M31 globular cluster properties and FUV photometry}
	\label{tab:data}
	\begin{tabular}{cccccccccccccc}
	\hline
		Name$^{a}$ &
        $g^{a}$ &
        $g-i^{a}$ &
        $M^{b}$ &
        $[Fe/H]^{c}$ &
        $E(B-V)^{c}$ &
        $r_{h}^{d}$ &
        $\tau_{h}^{d}$ &
        $\mu_{0}^{d}$ &
        \multicolumn{2}{c}{\galex$^{e}$} &
        $\hst^{f}$ \\
    	 &
         &
         &
        $10^{6}M_{\odot}$ &
         &
         &
        pc &
        Gyr &
        {\footnotesize mag/asec$^{2}$ } &
        $FUV$ &
        $NUV$ &
        $F140LP$ \\
    \hline
          B127-G185 & 14.16 & 1.21 & 4.3 & -0.80 & 0.09 & 2.56 & 2.3 & 17.32 & 20.66 $\pm$ 0.11 & 19.39 $\pm$ 0.08   & 20.36 $\pm$ 0.02 \\
          B131-G189 & 15.33 & 0.83 & 1.0 & -0.81 & 0.12 & 1.83 & 0.8 & 16.55 & 20.81 $\pm$ 0.13 & 19.89 $\pm$ 0.11   & 20.52 $\pm$ 0.03 \\
          B178-G229 & 14.91 & 0.73 & 1.4 & -1.51 & 0.12 & 2.20 & 1.1 & 17.49 & 19.48 $\pm$ 0.03 & 18.73 $\pm$ 0.02   & 19.27 $\pm$ 0.01 \\
          B179-G230 & 15.36 & 0.83 & 1.0 & -1.10 & 0.10 & 2.69 & 1.3 & 17.38 & 21.67 $\pm$ 0.14 & 19.91 $\pm$ 0.04   & 21.68 $\pm$ 0.03 \\
          B193-G244 & 15.36 & 1.09 & 1.3 & -0.44 & 0.11 & 2.03 & 1.0 & 16.93 & 21.96 $\pm$ 0.10 & 21.24 $\pm$ 0.05   & 23.12 $\pm$ 0.20 \\
          B205-G256 & 15.29 & 0.78 & 1.0 & -1.34 & 0.14 & 2.16 & 1.0 & 17.95 & 22.07 $\pm$ 0.19 & 20.10 $\pm$ 0.03   & 21.70 $\pm$ 0.03 \\
          B206-G257 & 14.94 & 0.75 & 1.4 & -1.45 & 0.13 & 2.81 & 1.6 & 17.70 & 20.43 $\pm$ 0.04 & 19.20 $\pm$ 0.02   & 20.09 $\pm$ 0.01 \\
          B218-G272 & 14.55 & 0.83 & 2.1 & -1.19 & 0.14 & 3.19 & 2.4 & 16.33 & 19.74 $\pm$ 0.05 & 19.07 $\pm$ 0.02   & 19.41 $\pm$ 0.01 \\
          B224-G279 & 15.10 & 0.66 & 1.1 & -1.80 & 0.13 & 6.95 & 5.8 & 20.26 & 20.78 $\pm$ 0.05 & 19.36 $\pm$ 0.02   & 20.46 $\pm$ 0.01 \\
          B225-G280 & 14.21 & 0.99 & 3.4 & -0.67 & 0.10 & 2.12 & 1.6 & 15.71 & 20.17 $\pm$ 0.05 & 19.33 $\pm$ 0.02   & 19.84 $\pm$ 0.01 \\
          B232-G286 & 15.46 & 0.67 & 0.8 & -1.83 & 0.14 & 2.43 & 1.0 & 17.25 & 20.67 $\pm$ 0.06 & 19.49 $\pm$ 0.02   & 20.34 $\pm$ 0.01 \\
          B472-D064 & 15.06 & 0.75 & 1.2 & -1.45 & 0.13 & 1.71 & 0.7 & 17.03 & 19.94 $\pm$ 0.04 & 19.03 $\pm$ 0.02   & 19.84 $\pm$ 0.03 \\
		\hline
	\end{tabular}

    \flushleft
$^{a}$Dereddened ``total" $g-$band magnitude and $g-i$ colour from SDSS photometry \citep{Peacock10a} \\
$^{b}$mass of the cluster inferred from $i$-band mag, assuming a distance modulus of 24.36 and a mass to light ratio $M/L_{i}=1.8$ \citep{Maraston05} \\
$^{c}$Metallicity and extinction \citep[][and references therein]{Fan08}. Higher resolution spectroscopy is only available for a few of these clusters, but we note that the authors caution significant systematic errors may exist in these low resolution metallicities \citep{Colucci09, Sakari16}. Throughout this paper, we utilize optical colour as a proxy for metallicity. \\
$^{d}$The half-light radius ($r_{h}$), half light relaxation time ($\tau_{h}$, from Equation \ref{eq:tau_h}) and central F336W surface brightness ($\mu_{0}$, see Section \ref{sec:sbp}).\\
$^{e}$``total" FUV and NUV magnitudes from \galex photometry \citep{Rey07}. \\
$^{f}$``total" FUV magnitudes (within a $5\arcsec$ aperture) from our F140LP observations. See Section \ref{sec:mag}. \\

\end{table*}

\section{M31's globular clusters}
\label{sec:sample}

M31 hosts the largest population of globular clusters in the Local Group \citep[$>400$, e.g.][]{Peacock10a}. At a distance of 744~kpc \citep{Vilardell10}, this cluster system is an important bridge between the well studied and spatially resolved stellar populations in the Galactic globular clusters and the unresolved, but far more numerous, extra-galactic globular cluster systems.

Globular clusters at the distance of M31 have median core radii, $r_{c}\sim 0.25\arcsec$ (0.8~pc) and half light radii, $r_{h}\sim 0.7\arcsec$ \citep[2.7~pc;][]{Barmby07, Peacock10a}. Observations using the \hst therefore resolve their integrated emission to scales similar to their core radii, while brighter stars outside of the cluster cores can also be resolved in such observations \citep[e.g.][]{Brown04, Perina12}.

This paper considers 12 of M31's globular clusters. All of these clusters have published integrated photometry in the SDSS $ugriz$ bands \citep{Peacock10a} and \galex FUV+NUV bands \citep{Rey07}. These clusters are all quite bright and massive with $i<14.8$~mag, which corresponds to mass,~$M>0.8\times 10^{6}M_{\odot}$ (assuming $i_{\odot}=4.58$~mag and $M/L_{i} = 1.8$, based on a 12~Gyr simple stellar population with $[Z/H]=-1.35$ and a Kroupa IMF; \citealt{Maraston05}). The clusters are also known to be bright in the FUV \citep[$FUV<22.1$~mag;][]{Rey09}. Additionally, all of the clusters have relatively low extinction, with $E(B-V)<0.16$ \citep[][]{Fan08}. This minimizes errors due to uncertainty in the extinction curve in the ultraviolet, which we take to have $R_{FUV}=8.2$ in the FUV (based on \citealt{Cardelli89} and as used by \citealt{Rey07}).

Structural parameters have been published for most of these clusters from ground based $K-$band observations \citep{Peacock10a} and \hst ACS or WFPC2 observations \citep{Barmby07, Strader11}. For the remaining two clusters, we fit their F336W images using the {\sc ishape} code \citep{Larsen99}. For each cluster, we calculate the half-light relaxation time via \citep{Binney87,McLaughlin05}:

\begin{equation} \label{eq:tau_h}
 \tau_{h} = 2.06 \times \frac{10^{6}}{ln(0.4M_{tot}/m_{\star})} \frac{M_{tot}^{0.5}}{m_{\star}} r_{h}^{1.5}
\end{equation}
\\
Where $m_{\star}$ is the average stellar mass. For all clusters, we take $m_{\star}=0.5M_{\odot}$. Table \ref{tab:data} lists our sample of globular clusters, their structural parameters, and integrated photometry. The final column shows the total FUV magnitudes derived from this study (See Section \ref{sec:mag}).

\begin{figure*}
\centering
\includegraphics[width=0.75\textwidth]{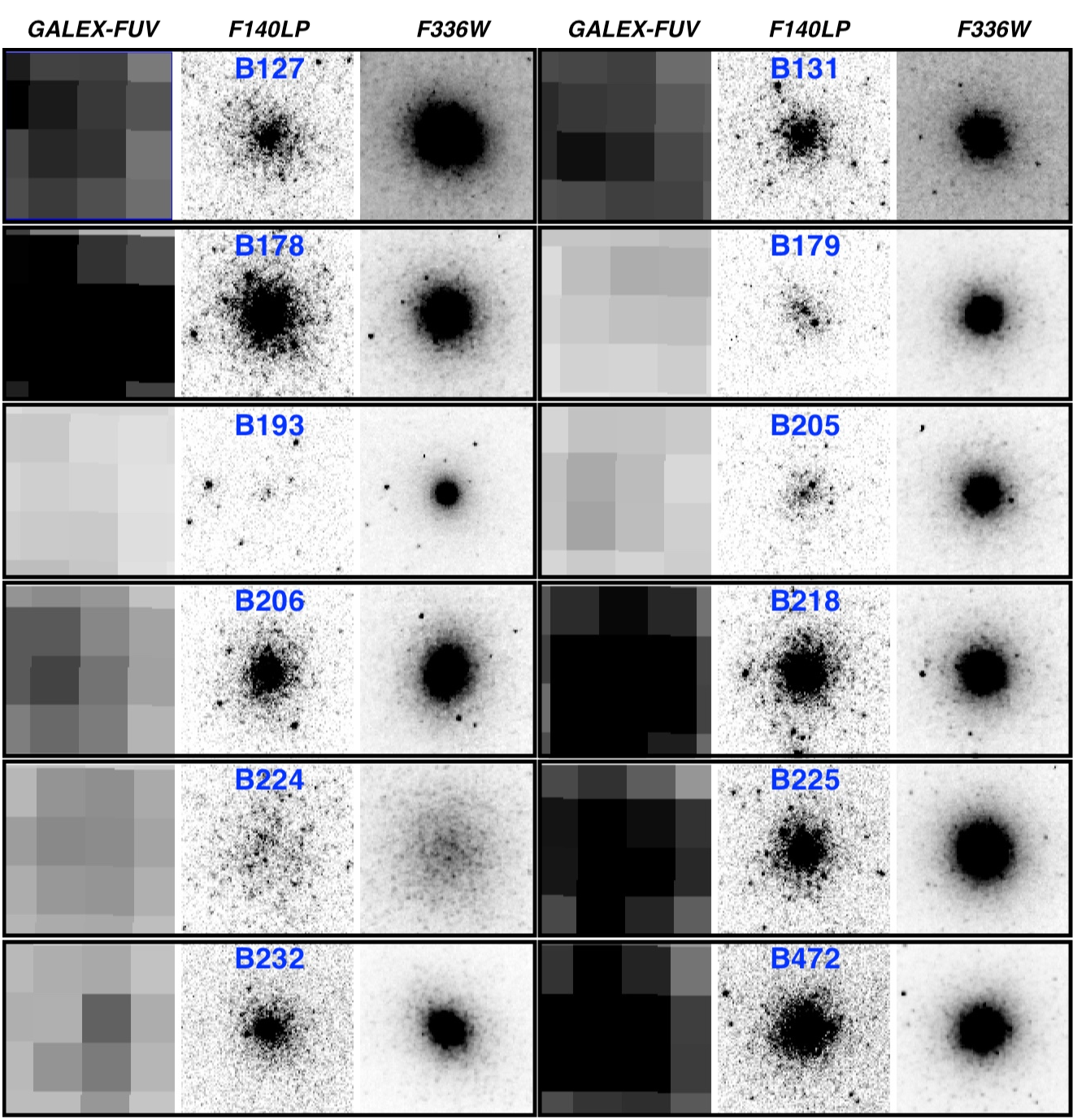}
\caption{\label{fig:thumb} Thumbnails of the 12 globular clusters studied. From top left to bottom right, these are B127, B131, B178, B179, B193, B205, B206, B218, B224, B225, B232, B472, respectively.
For each cluster we show observations from \galex FUV, \hst ACS/SBC F140LP, and \hst WFC3/UVIS F336W, left to right, respectively. Each image is $5\arcsec\times5\arcsec$. }
\end{figure*}

\section{Data and Analysis}
\label{sec:data}

\subsection{High spatial resolution HST observations}
\label{sec:fuv_data}

To investigate the hot populations in these globular clusters, we obtained \hst ACS/SBC observations through the F140LP filter under program ID 14132 (PI Peacock). These observations were taken in September and October of 2016 with each cluster observed for one orbit using a four point dither pattern; resulting in total exposure times for each cluster of 2713s. The F140LP filter has a pivot wavelength of 1528\AA \ and a similar transmission profile to the \galex FUV filter. We utilize the pipeline reduced products for each exposure, the `flt' files. For each cluster we combine the four exposures together using the {\sc drizzlepac} task {\sc astrodrizzle}, run under {\sc astroconda}. We drizzle combine onto a 0.025\arcsec grid with $`final\_pixfrac'=1.0$.

Multi-wavelength \hst observations of these clusters are also available at redder wavelengths from the Panchromatic Hubble Andromeda Treasury (PHAT) survey \citep{Dalcanton12a}. This provides near-UV WFC3/UVIS F275W + F336W observations and optical ACS/WFC F475W + F814W observations. We utilize the reduced and mosaicked WFC3/UVIS F336W ``brick" images provided by this survey\footnote{Obtained from https://archive.stsci.edu/prepds/phat/datalist.html}. These images have a pixel scale of $0.04\arcsec$. For details of the reduction of these PHAT data see \citet{Williams14}. For each cluster, we cut sections from these bricks which cover the ACS/SBC fields and align them using the {\sc drizzlepac} task {\sc tweakreg}.

Thumbnails of these ACS/SBC/F140LP (FUV) and WFC3/UVIS/F336W (similar to a $u$-band filter) images are presented in Figure \ref{fig:thumb}. We also show the clusters  from \galex FUV observations, which have a similar wavelength coverage to the F140LP filter, but a much lower spatial resolution. It can be seen that all of the clusters are well resolved in these \hst images, with many of their brighter stars also resolved.

\subsection{Integrated FUV photometry}
\label{sec:mag}

We perform photometry on these images using the {\sc daophot/phot} task, implemented under {\sc pyraf}. The flux is measure through a range of log-spaced apertures with radii, $0.05<r<5.0\arcsec$, which corresponds to $0.2<r<18.0$~pc at a distance of 744~kpc. The cluster centres were defined based on the F336W images of the clusters. Due to the higher stochasticity in the F140LP profiles we fixed the cluster centres based on their locations in the aligned F336W images. We note that consistent profiles were also obtained using the {\sc stsdas} task {\sc ellipse}, but this task assumes a smooth positive gradient towards the cluster centre and failed to fit isophotes to the inner regions for some of the clusters, we therefore chose to use the simple aperture photometry instead.

We consider several different techniques and spatial regions for estimating the background level. The background in the F140LP images is found to be dominated by discrete point sources. We therefore choose to estimate the background level for all apertures as the mean value within a fixed sky annulus of $5-7\arcsec$, with no sigma clipping.

The instrumental magnitudes are placed on the ABMAG system using zeropoints calculated from the calibrations provided by the \hst/ACS pipeline using the relation:

\begin{equation}
\begin{split}
m_{0,AB} = -2.5\ &log_{10}(PHOTFLAM) -21.10 \\
					&-5.0\ log_{10}(PHOTPLAM) + 18.6921
\end{split}
\end{equation}
\\
For these F140LP data, $PHOTFLAM = 2.7128664\times 10^{-17}$ and $PHOTPLAM = 1528.0$.

In addition to total magnitudes within each aperture, we also calculate the average surface brightness within each annulus as the difference in flux between successive apertures:

\begin{equation}
\mu_{i} = m_{0,AB} - 2.5 log_{10} \left( \frac{ F_{i}-F_{i-1} } {A_{i}} \right) \\
\end{equation}
\\
Here, $\mu_{i}$ is the surface brightness in annulus $i$ (in mag/asec$^{2}$), $F_{i}$ is the flux within aperture $i$ and $A_{i}$ is the area of the annulus (in asec$^{2}$).

\begin{figure}
\centering
\includegraphics[width=0.48\textwidth]{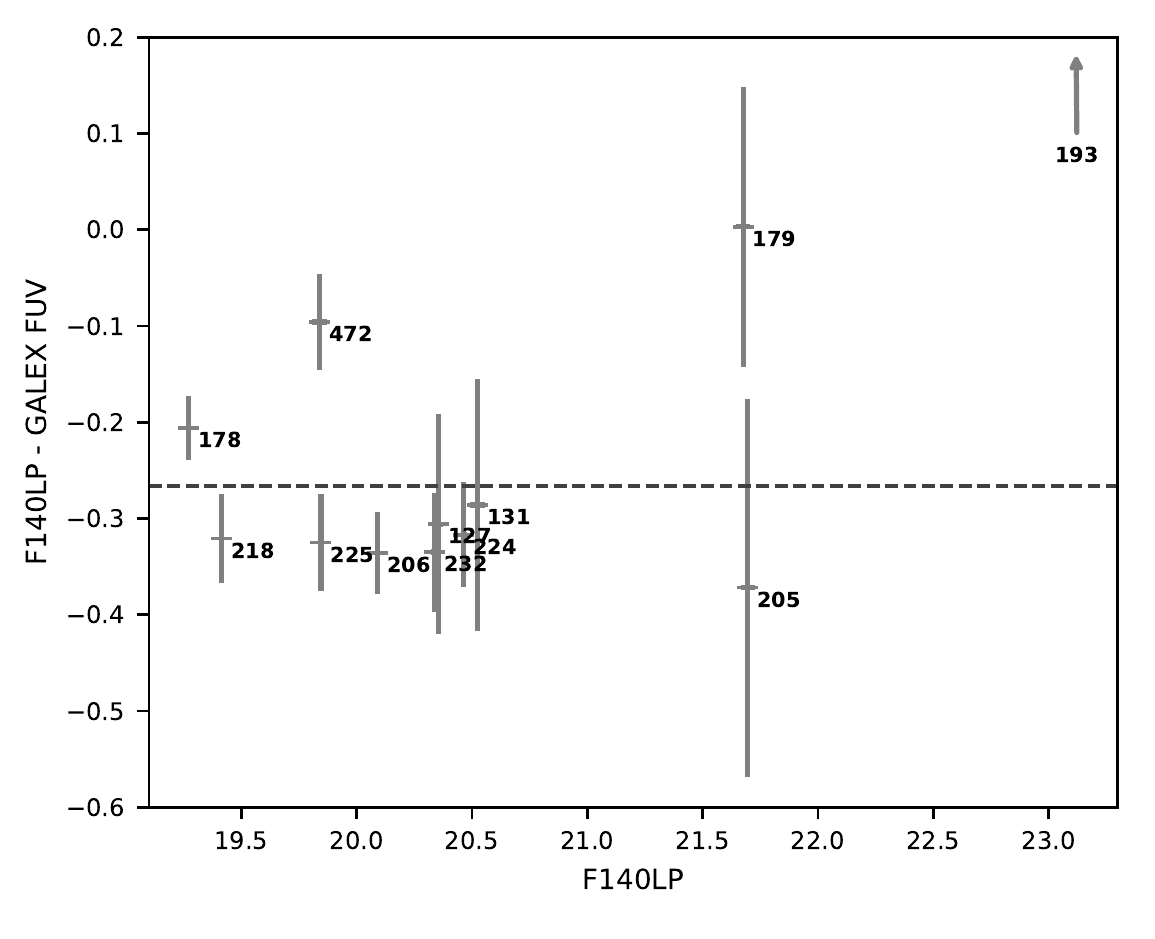}
\caption{\label{fig:sbc_galex}A comparison of the FUV magnitudes of M31's globular clusters from published \galex $FUV$ photometry and our \hst ACS/SBC $F140LP$ photometry.}
\end{figure}

In Figure \ref{fig:sbc_galex}, we show the FUV magnitude of these clusters within a $5\arcsec$ aperture. This aperture corresponds to a physical scale of 18~pc and, while cluster stars may be expected beyond this radius,  the surface brightness profiles are found to be consistent with the sky level at this distance. We compare these magnitudes to \galex photometry of the clusters, as published by \citet{Rey07}. Because of the large \galex PSF, these clusters are consistent with point sources in such observations and their photometry was obtained through a 4.5$\arcsec$ aperture with an aperture correction derived from bright stars in the image.

For 11 of the 12 clusters observed our \hst/SBC/F140LP photometry is consistent with a slight offset from the \galex FUV photometry, with $F140LP=FUV-0.27$. Some of this offset may result from either differences in the estimation of the background level or from the aperture correction applied to the \galex observations. However, the clusters may also be slightly brighter in these F140LP observations due to a red leak \citep[which is known to affect ACS/SBC observations;][]{Boffi08}. In \citet{Peacock17a}, a small red-leak was also suggested in FUV photometry of M~87's globular clusters, based on comparing observations obtained using the similarly designed \hst STIS MAMA detector with the alternate technology and independently calibrated WFPC2 detector. We therefore caution that $\sim 20\%$ of the FUV flux in these F140LP observations may come from redder wavelengths. We propose that such a correction should be considered when analysing \hst ACS/SBC or STIS FUV observations of globular clusters. Ideally the now recommended `dual filter' approach can be used to directly identify and remove such red leak (such as the recent observations of clusters in NGC~3115; Proposal ID 14738, PI Kundu).

These F140LP observations do confirm significant FUV emission from these clusters. Additionally, we find no evidence for a trend with luminosity when comparing to the \galex observations, as might be expected if contamination was effecting the \galex observations. We therefore confirm that the published \galex integrated photometry of these clusters is likely reliable, despite the low spatial resolution and signal to noise of the observations. This includes two metal-rich clusters (B127 and B225) which are much more luminous in the FUV than predicted by canonical models for HB star morphology \citep[see][and Section \ref{sec:ehb_msp}]{Rey07, Peacock17a}.

One cluster (B193) is found to be significantly fainter in our observations. As can be seen in Figure \ref{fig:thumb}, this cluster is extremely faint in these F140LP observations. We note that this source is a low significance detection in the \galex photometry and believe its published photometry may be unreliable due to either contamination from nearby sources or errors in the background estimation.

\subsection{Radial profiles in the ultraviolet: FUV- and u-band}
\label{sec:sbp}

\begin{figure*}
\centering
\includegraphics[width=0.9\textwidth]{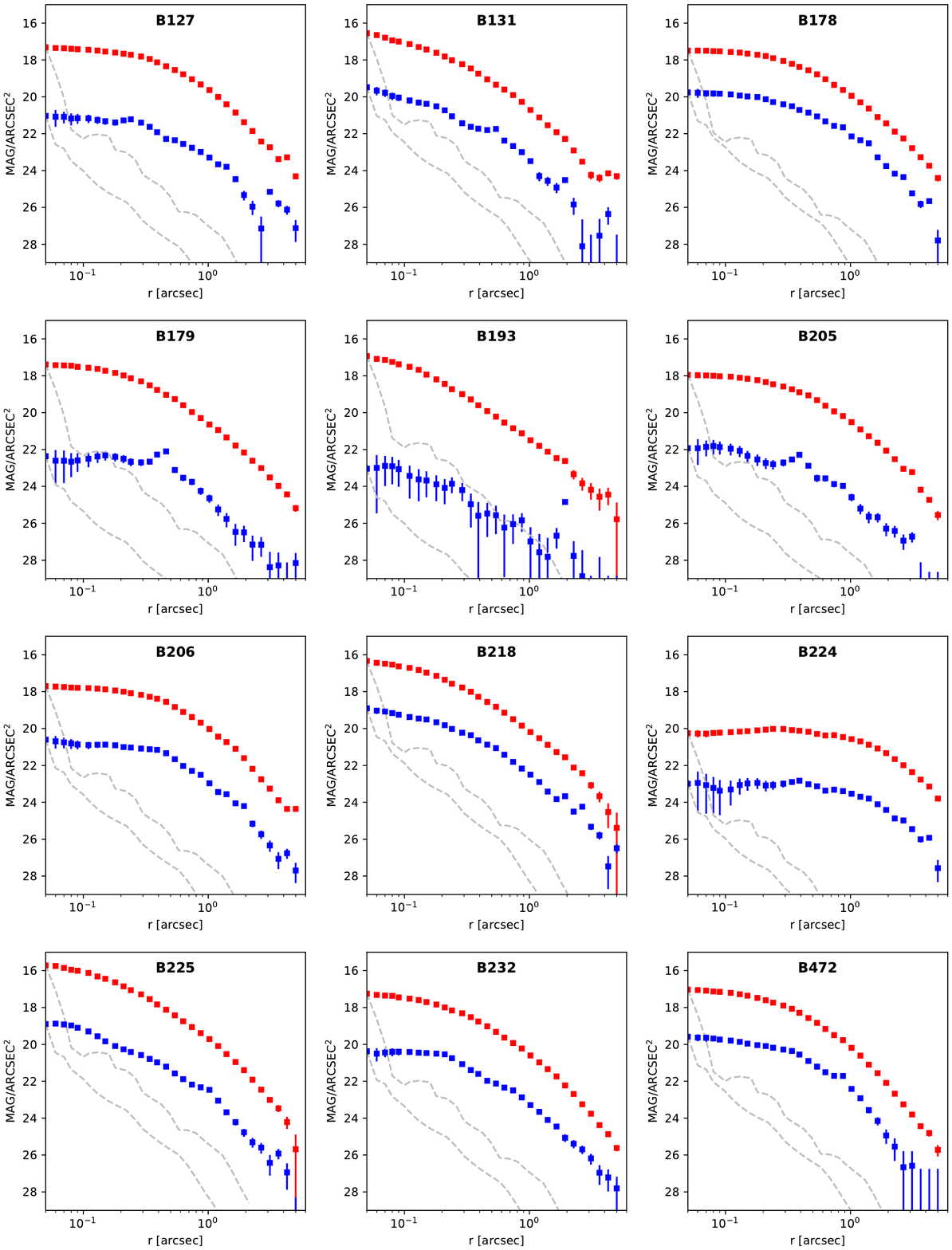}
\caption{\label{fig:profs} Surface brightness profiles ($mag/arcsec^{2}$) of the 12 globular clusters through the F140LP (blue points) and F336W (red points) filters. Grey dashed lines show the expected profiles of a point source, based on the TinyTim simulator and scaled to have the same surface brightness as the clusters at $0.05\arcsec$.}
\end{figure*}

\begin{figure*}
\centering
\includegraphics[width=0.9\textwidth]{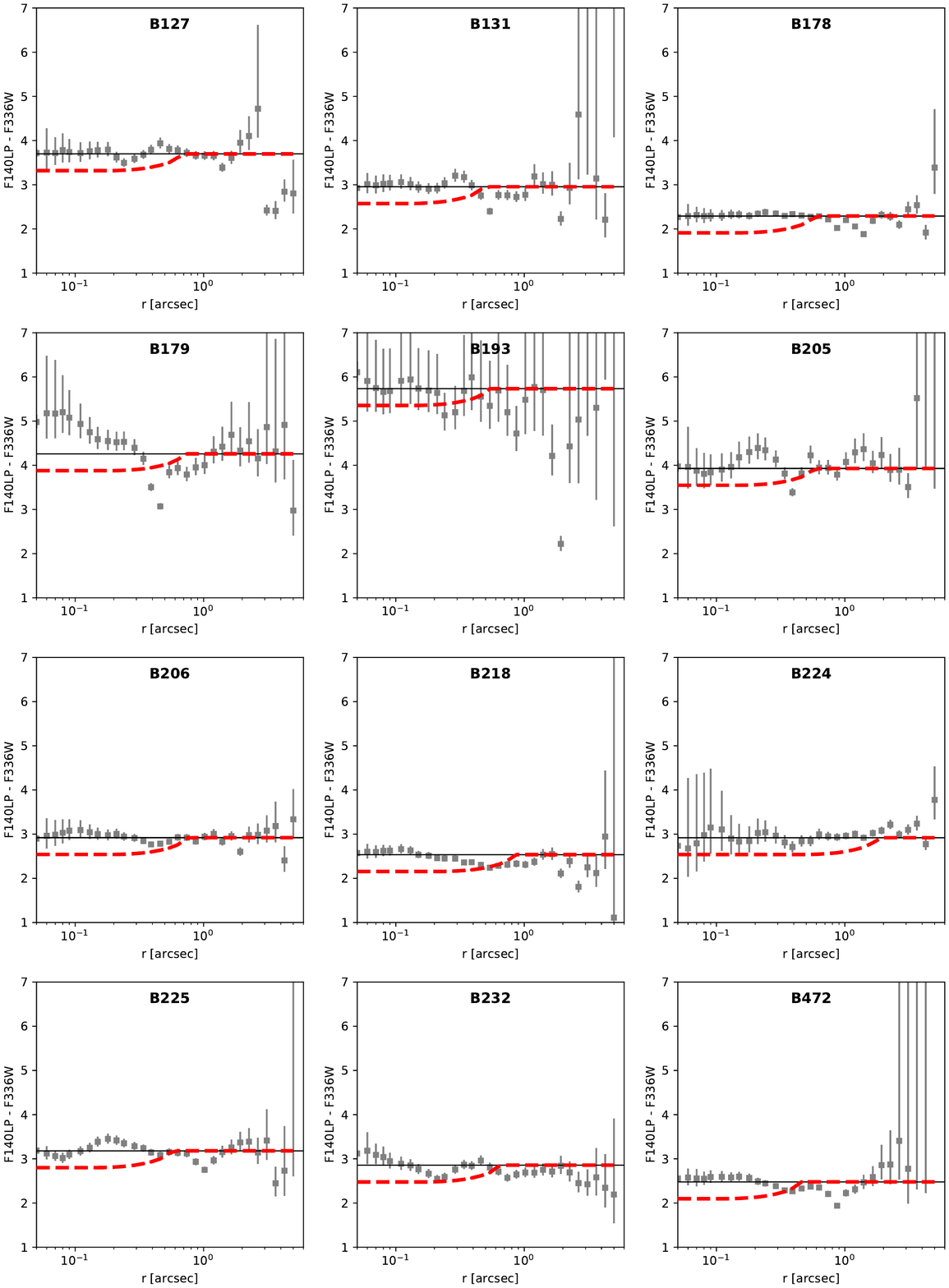}
\caption{\label{fig:color_profs} $F140LP - F336W$ ($FUV-u$) profiles of 12 of M31's old globular clusters. The solid black line shows the weighted mean colour out to $5\arcsec$. The red-dashed line shows the predicted profiles if the clusters host centrally concentrated second populations of Helium enhanced stars, similar to those observed in the Galactic globular clusters by \citet[][see Section \ref{sec:msp_radial}]{Lardo11}. }
\end{figure*}

Figure \ref{fig:profs} shows the surface brightness profiles of our sample of 12 clusters in F140LP (FUV, blue points) and F336W ($u$, red points). For comparison with the cluster profiles, the dashed-grey lines in Figure \ref{fig:profs} show the profile expected for a point source. This is estimated using the TinyTim simulations \citep{Krist11}\footnote{http://tinytim.stsci.edu/cgi-bin/tinytimweb.cgi}. Profiles are constructed from the resulting PSF images in a similar fashion to the cluster images and scaled to match the cluster surface brightness at $0.05\arcsec$. It can be seen that all of the clusters are well resolved.

The F140LP profiles reflect some stochasticity from the fact that the light is dominated by less numerous blue and extreme HB stars, while redder bands, including F336W, are dominated by much more numerous, cooler stars. Detailed comparison confirms that the ripples in the profiles are produced by individual FUV bright stars. We choose not to subtract these stars from the profiles since cluster stars could be this bright and the magnitude to which we remove FUV bright stars would be arbitrary.

With the exception of features associated with individual FUV bright point sources, the profiles of the clusters are found to be quite similar at F140LP and F336W wavelengths. To better consider variations in the profiles, we show in Figure \ref{fig:color_profs} the $F140LP-F336W$ profiles of the clusters. These figures include the weighted mean colour of clusters over the full radial region considered ($r<5\arcsec$; solid black line). The average $FUV-u$ colours of the clusters are found to vary by 2.5 magnitudes. Some variation is observed in the colour with radius. This can generally be attributed to individual bright sources. However, no significant trends in the colours of these clusters with radius are observed.

\subsection{Distribution of blue/ extreme HB stars}
\label{sec:radial_hb}

The $F140LP-F336W$ colour probes the HB star morphology. This is because the presence of very hot blue/extreme HB stars will strongly influence the FUV flux, but will have far less affect on redder bands (including F336W) which will be dominated by emission from more common cooler stars.  Therefore, these surface brightness profiles in F140LP and F336W can be used to test whether the hot populations in these clusters are spatially distinct. We identify no significant radial trends in the $F140LP-F336W$ colours of these clusters. These data therefore suggest that the blue/extreme HB stars are well mixed with the other stellar populations.

Similar FUV profiles have not been published for the Galactic globular clusters. However, the distribution of the resolved HB populations have been studied in some of these clusters \citep[e.g.][]{Iannicola09, Vanderbeke15}. \citet{Vanderbeke15} recently traced the radial distribution of red, blue and extreme HB stars in 48 Galactic globular clusters, by identifying the populations in ground based optical $g-$ and $z-$band photometry. This work is sensitive to a similar radial range to our surface brightness profiles, from around the core/half-light radius to a few half light radii. \citet{Vanderbeke15} drew similar conclusions to our surface brightness profiles of M~31's globular clusters, with 80$\%$ of their Galactic globular cluster sample having similarly distributed HB stars. Some of the remaining clusters did show centrally concentrated extreme HB populations, while they were less concentrated in three of the clusters.

Below we discus possible implications and origins of the well mixed FUV bright populations observed in M~31's clusters.

\begin{figure}
\centering
\includegraphics[angle=270, width=0.48\textwidth]{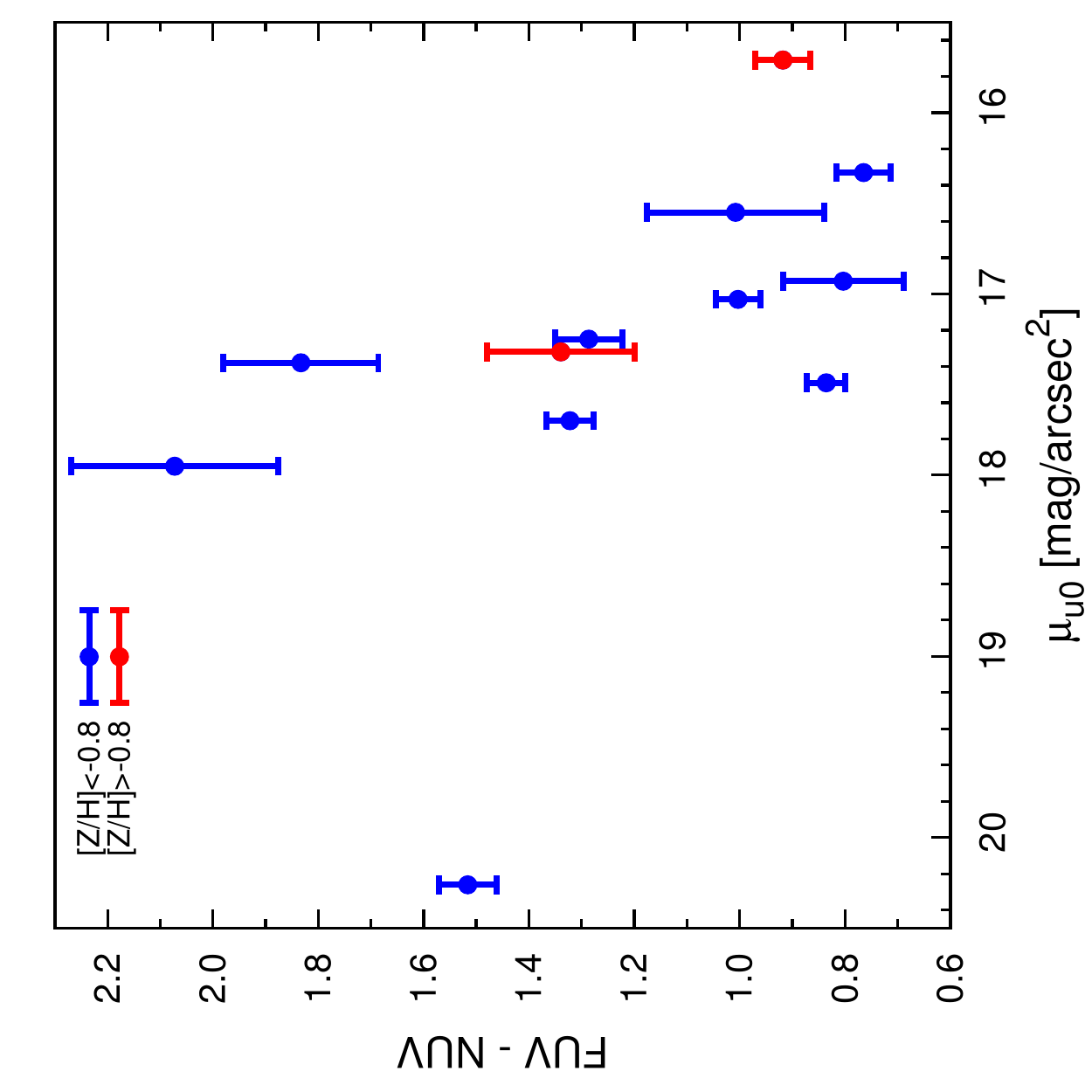}
\caption{\label{fig:density} The integrated \galex $FUV-NUV$ colour of our sample of M31's clusters as a function of central stellar density. Metal-rich ($[Z/H]>-0.8$) and metal-poor clusters are identified as red and blue points, respectively. We note that B193 is excluded from this plot, since we believe its \galex colours may be unreliable (see Section \ref{sec:mag}). }
\end{figure}

\begin{figure}
\centering
\includegraphics[angle=270, width=0.48\textwidth]{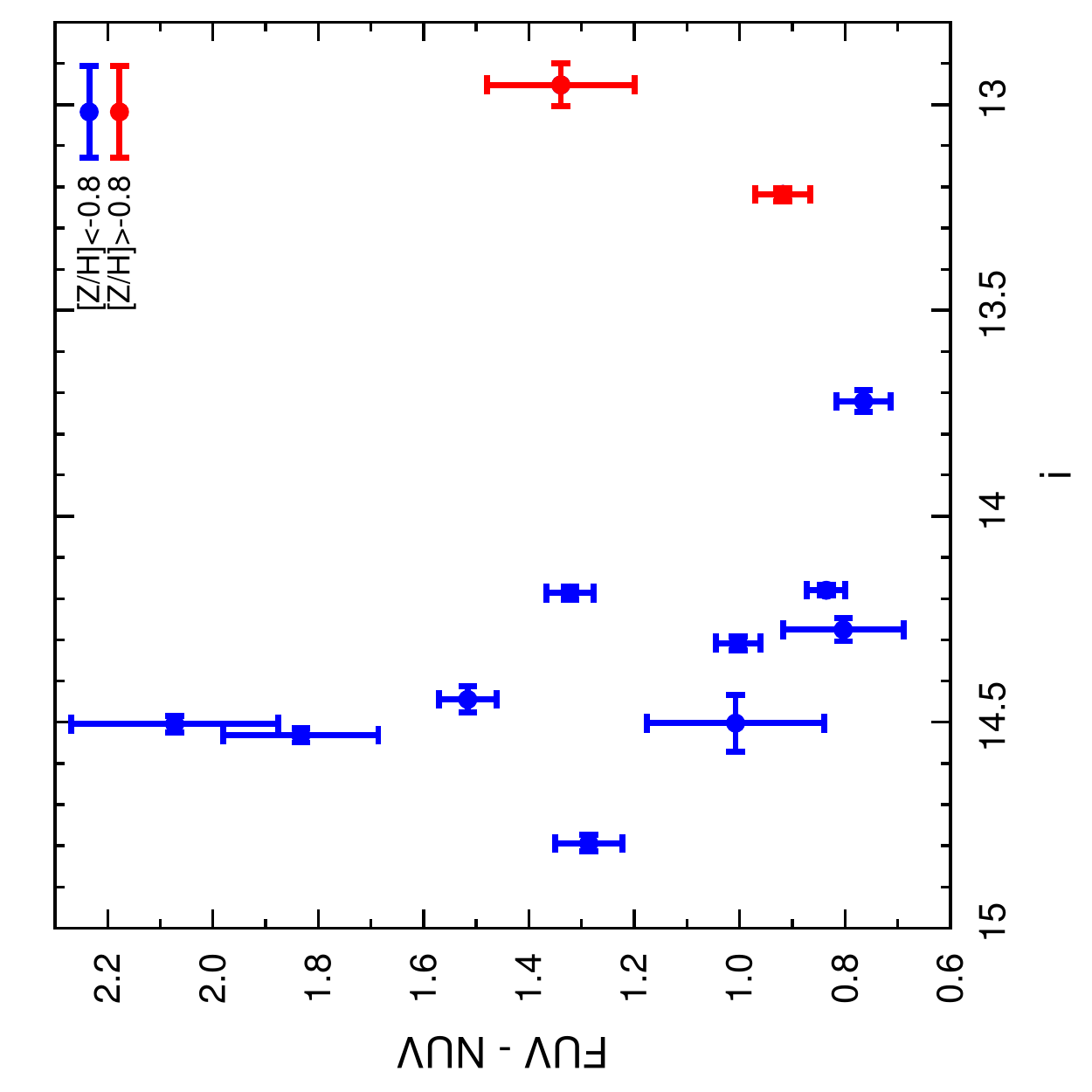}
\caption{\label{fig:mass} Same as Figure \ref{fig:density}, but comparing $FUV-NUV$ to total $i$-band magnitude. }
\end{figure}

\section{Dynamical formation of extreme HB stars}
\label{sec:ehb_dyn}

Dynamical interactions between stars in the dense cores of globular clusters may enhance the formation of extreme-HB stars. Several processes have been proposed to produce new or hotter HB stars in such dense environments, including: helium dredge up in close interactions; enhanced mass loss in tightened binary systems; and the merger of two white dwarfs to produce an extreme-HB like star (as discussed in Section \ref{sec:intro}).

Some observations of the Galactic globular clusters suggest that dynamical interactions may produce bluer HB stars, with measures of the clusters' HB temperatures correlating with the stellar density \citep[e.g.][]{Fusi_Pecci93,Buonanno97,Rich97,Dotter10}. However, other studies have proposed a lack of correlation \citep[e.g.][]{VandenBergh93, Recio_blanco06, Dalessandro12}. In M31's clusters, it has previously been shown that their $FUV-NUV$ colour correlates with stellar density, with denser clusters having bluer $FUV-NUV$ \citep{Peacock11b}. In Figure \ref{fig:density}, we show that this trend is reproduced by our sample of clusters with $FUV-NUV$ correlating with central $u-$band surface brightness. We also show $FUV-NUV$ as a function of the total $i$-band magnitude of the clusters (Figure \ref{fig:mass}). This shows a similar trend, which is unsurprising, given that cluster mass is known to correlate with central stellar density \citep[e.g.][]{McLaughlin05, Peacock10b}. One explanation for this correlation is that denser and more massive clusters are brighter in the FUV due to increased dynamical interactions producing bluer HB stars and hence more FUV emission.

A key signature of a dynamically enhanced population is it should be centrally concentrated -- since it will be preferentially produced in the cluster cores, where the stellar density is highest. This is observed for another population of dynamically enhanced systems, low mass X-ray binaries. These systems are generally found in the core regions of Galactic globular clusters and do preferentially reside in denser clusters with higher interaction rates \citep{Bellazzini95, Pooley03, Peacock09}. It is also worth noting that the lifetimes of HB stars are typically an order of magnitude less than the half light relaxation times of these clusters. It is therefore unlikely centrally formed extreme HB stars have become significantly mixed with the other populations.

None of the clusters in our sample are observed to have significantly bluer cores (see Figure \ref{fig:color_profs}). We therefore conclude that there is no evidence that the excess FUV emission from these dense clusters is the result of dynamical interactions  producing significant populations of extreme-HB stars. We note that this is despite our sample including clusters with very high central densities and hence high stellar interaction rates (see Table \ref{tab:data}). Indeed, the presence of LMXBs in two of these clusters confirms that stellar interactions are occurring \citep[B193 and B225;][]{Trudolyubov04, Peacock10b}. An alternate explanation for these FUV bright clusters is therefore preferred, such as sub-populations of helium enriched stars (discussed in Section \ref{sec:ehb_msp}, below).

\section{Second populations in M~31's globular clusters}
\label{sec:ehb_msp}

\begin{figure*}
\centering
\includegraphics[angle=270, width=1.00\textwidth]{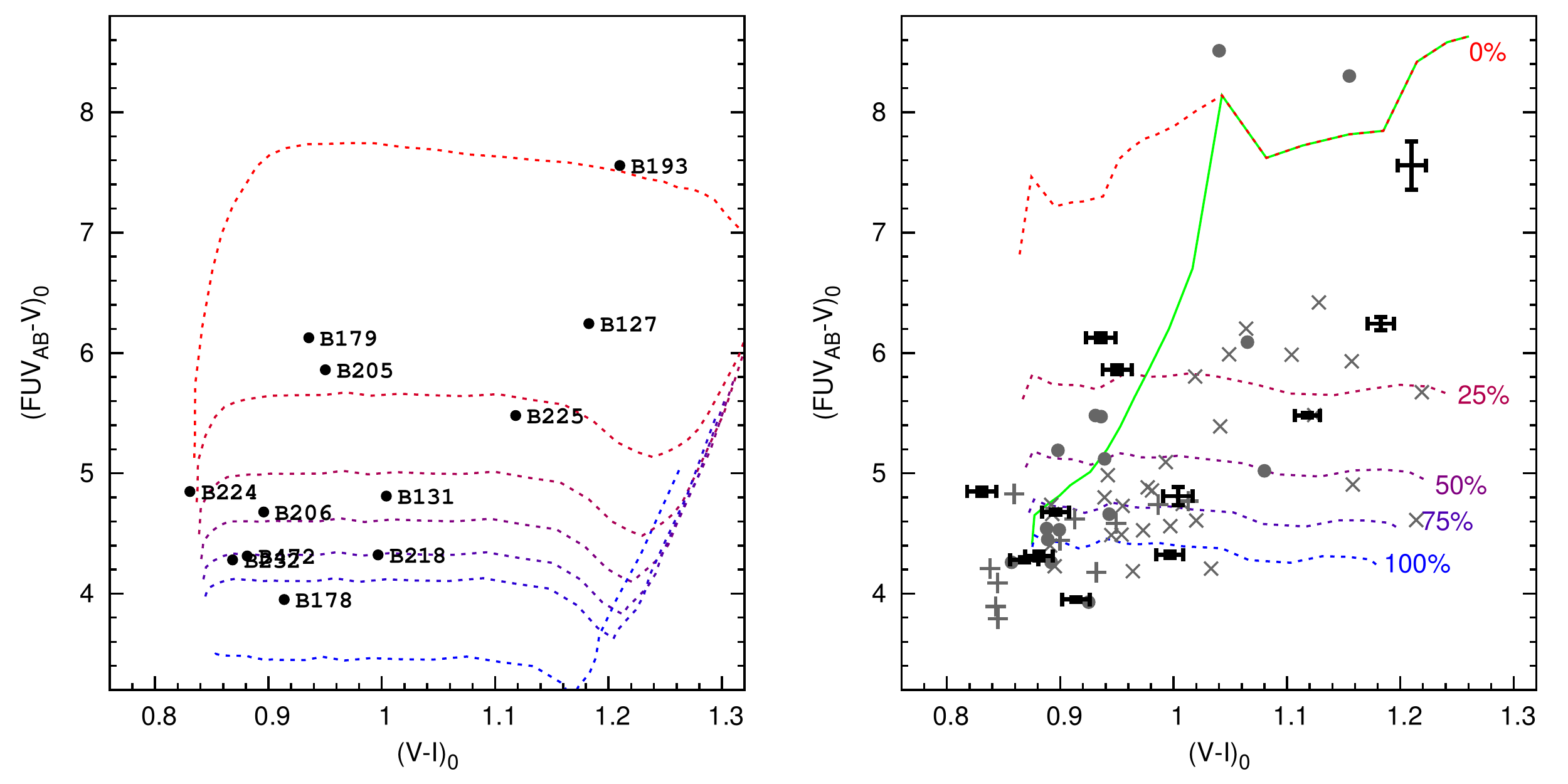}
\caption{\label{fig:msp}Integrated $FUV-V$ of M31's clusters as a function of their $V-I$ optical colour (labelled points in the left panel and black points with errorbars in the right panel). In the right panel, we also show clusters in the Milky Way (grey points), other clusters in M31 (grey pluses) and M~87 (grey crosses). Only quite massive clusters (those with $M_{V}<-9.0$) are included, this reduces stochastic effects in the $FUV-V$ colour. The left panel shows predicted colours from the YPES models of \citet{Chung17}. These are for a stellar population that is composed of a primordial population with helium fraction, Y=0.23, and a second helium enhanced population with Y=0.38 (where the fraction of second population stars is $0,10,20,30,40,50,100\%$, red to blue lines). The right panel compares these data to the FSPS models of \citet{Conroy09} where the fractions of blue- to red-HB stars is $0,25,50,75,100\%$, red to blue lines. The green lines shows the variation predicted due to the influence of metallicity on the HB star population. }
\end{figure*}

It is now well established that the Galactic globular clusters host multiple stellar populations. While a comprehensive model for the formation of multiple populations in globular clusters is still lacking, most theories require hydrogen burning at high temperatures in order to reproduce the observed light element abundance variations. Therefore, second population stars are expected to be helium enhanced.

These different helium abundances can produce the observed split/broadened main sequences and red giant branches. However, their most dramatic effect is on the HB star morphology, since helium enhanced populations are expected to produce HB stars with much higher surface temperatures. It has been shown that the HB star morphologies of some Galactic globular clusters can be explained by including helium enhanced populations \citep[e.g.][]{Lee05, DAntona05, Joo13}. Further evidence comes from a correlation between size of the spread in the light element abundance of second population stars in a cluster with the maximum temperature of its HB stars \citep{Carretta10}, while \citet{Marino11} observed the Na and O abundances of red and blue HB stars in M4 and found that that they were consistent with those of the first and second generations, respectively.

The bluer HB stars produced by helium enriched second populations provide a signature for detecting second populations in extra-galactic globular clusters. This is because a population of blue/extreme-HB stars will dominate the integrated FUV emission from a cluster. Interestingly, \citet{Peacock17a} showed that extragalactic globular clusters in M~31 \citep{Rey07}, M~81 \citep{Mayya13}, and M~87 \citep{Sohn06, Peacock17a} often have bluer $FUV-V$ colours than expected for simple stellar populations models with standard helium abundances.

In Figure \ref{fig:msp}, we plot the FUV and optical colours of the M31 clusters studied here along with published FUV and optical colours of other clusters. Specifically, we include published FUV colours of clusters in the Milky Way \citep{Sohn06, Dalessandro12}, M~87 \citep{Sohn06, Peacock17a} and other clusters in M31 \citep[from \galex observations;][]{Rey07}. We limit this sample to only include quite massive clusters, those with $M_{V}<-9.0$. This ensures that the clusters have relatively large populations which reduces stochastic effects in the FUV colours. For details of these extragalactic cluster data see \citet[][and references therein]{Peacock17a}.

To interpret these cluster colours, we compare to predictions based on two different stellar populations synthesis models. We note that there is significant uncertainty in modelling the properties of the HB stars in a stellar population and these models are intended as a guide to the effect of including second populations of helium enhanced stars, rather than a precise comparison. The left panel shows the YPES models of \citet{Chung17} for an 11~Gyr population with metallicity in the range $-2.6 < [Z/H] < 0.6$ for clusters hosting two populations of stars; one with primordial helium abundances ($Y=0.23$) and the other with enhanced helium abundances ($Y=0.38$). The fraction of the helium enhanced second population is 0, 10, 20, 30, 40, 50, $100\%$ (red to blue lines, respectively). The right panel shows models from the FSPS models \citep{Conroy09,Conroy10}. These are based on the PADOVA stellar isochrones and BASEL stellar libraries for a 12~Gyr simple stellar population with metallicity in the range $-2.0 < [Z/H] < 0.0$. We plot models with different fractions of blue- to red-HB stars: 0, 25, 50, 75, $100\%$, red to blue lines, respectively. The green solid line in Figure \ref{fig:msp} shows the predicted variation due to metallicity; where the population varies from having only blue-HB stars at low metallicity to only red-HB stars at high metallicity.

The predicted cluster colours vary due to both the different models used and the details of the populations assumed. However, the general trend is very similar and consistent with the empirical conclusions based on the Milky Way's clusters; that increasing the fraction of helium enhanced second population stars produces bluer HB stars and hence bluer $FUV-V$ colours \citep[see also the models considered in][for similar conclusions]{Rey07}.

Metallicity is known to influence the HB population of clusters, with metal-rich globular clusters producing red HB stars for standard helium abundances. These relatively cool HB stars should result in very red $FUV-V$ colours, such as the Galactic globular cluster 47 Tuc and the cluster B193 from this study. However, two of the metal rich clusters in our sample (B127 and B225) have bluer colours than predicted based on their metallicity, suggesting the presence of a significant fraction of helium enhanced second population stars. These clusters may be analogous to the metal-rich Galactic globular cluster NGC~6441, which has both red-HB stars and a tail of much hotter HB stars resulting in its relatively bright FUV emission. We note that similarly $FUV$ bright metal-rich clusters are also observed in the Milky Way (NGC~6388 and NGC~6441), M~87 \citep{Sohn06, Peacock17a} and M~81 \citep[GC1;][]{Mayya13}.

Metal-poor globular clusters tend to host blue HB populations for standard helium abundances. This means that the presence of second populations of hotter helium enhanced stars has less of an influence on their integrated $FUV-V$ colour. However, significant variation is observed at a given $V-I$ (or metallicity) with the clusters B131, B178 and B218 having relatively blue $FUV-V$ for their metallicity -- these could be explained by the presence of higher fractions (or more helium enhanced) second populations with hotter HB stars. The integrated colours of some of the clusters in our sample are therefore consistent with the presence of second populations of helium enhanced stars, similar to those observed in the Milky Way.

\citet{Perina12} have previously studied the HB populations of M31's outer halo globular clusters based on optical HST colour magnitude diagrams. Interestingly, they observed that their sample of clusters have slightly redder HB morphology for a given metallcity than the Galactic clusters. A possible explanation is that these clusters are slightly younger than the Galactic comparison sample, potentially consistent with their location in the outer halo and substructure association. Such work represents a complimentary approach to integrated FUV observations. The resolved CMDs allow direct counts of stars redder and bluer than the instability strip, but are less sensitive to the hottest HB stars. Similar to our conclusions, \citet{Perina12} showed that integrated FUV emission from some of their sample are consistent with extreme-HB stars and that the ratio of red to blue HB stars in some metal rich clusters (G1 and B045) may suggest the presence of multiple populations.

\subsection{A correlation between the mass/density of M31's clusters and their second populations?}
\label{sec:msp_mass}

As discussed in Section \ref{sec:ehb_dyn}, M31's clusters show a trend such that more massive, denser clusters are bluer in $FUV-NUV$ (see Figures \ref{fig:density} and \ref{fig:mass}). The $FUV-NUV$ colour is sensitive to the temperature of the HB star populations. Second populations in these clusters could explain this trend, if their fraction or helium abundance correlates with cluster mass (and hence density). This would produce hotter HB stars in denser clusters and hence the bluer ultraviolet colour. Interestingly, such a trend has been proposed in the Milky Way's globular clusters, where the helium enhancement of the second population appears to correlate with cluster mass \citep[e.g.][]{Carretta10, Milone17}. The trends observed in Figures \ref{fig:density} and \ref{fig:mass} can therefore be interpreted as evidence that M31's clusters host similar second populations of stars to the Milky Way, with their helium enhancement (and hence HB star temperatures) correlating with the mass and density of the clusters.

\subsection{Spatially distinct second populations}
\label{sec:msp_radial}

Some theories for the formation of multiple populations produce the second population centrally within the clusters. This is advantageous because, as the cluster evolves, it preferentially loses stars from the less concentrated first population -- hence reducing the mass budget problem. Simulations have shown that, if the relaxation times of the clusters are long enough, these populations can remain spatially distinct for a Hubble time \citep[e.g.][]{Vesperini13}. Therefore, identifying whether multiple populations are spatially distinct can place important constraints for theories of their formation.

Several independent studies have observed centrally concentrated second populations in the Galactic globular clusters. \citet{Lardo11} used Sloan digital sky survey observations of red giant branch stars to identify centrally concentrated second populations in seven out of the nine clusters they studied (over the radial range $0.5 r_{h} < r < 5 r_{h}$). Observations of individual clusters have drawn similar conclusions \citep[e.g.][]{Carretta10b, Kravtsov11, Beccari13, Simioni16}. However, there are some exceptions with evidence for well mixed populations in some clusters \citep{Dalessandro14, Vanderbeke15} and even less concentrated second populations \citep{Lim16,Dalessandro18}. Additionally, studies utilizing the HB star temperature as a tracer of first and second population stars found the populations to be well mixed in many clusters \citep{Iannicola09, Vanderbeke15}. If present, the lack of radial signatures identified in some of the Galactic clusters may be due to observing different radial regions (and associated relaxation times) or the sensitivity of different techniques for identifying the populations. Additionally, we note that cluster to cluster variations are expected due to differing relaxation times and known differences in both the fractions and abundances of their second populations.

\subsubsection{Testing for spatially distinct populations in M31s clusters}

The $FUV$ profiles of M31's clusters -- particularly those whose FUV colours suggest the presence of multiple populations -- can potentially extend this work to clusters beyond the Milky Way. If the second population is centrally concentrated, one would expect `excess' $FUV$ emission from the bluer second population HB stars to be more concentrated than the emission at redder wavelengths (which will probe the combined population).

Modelling the potential effects on the integrated $FUV-u$ profiles is challenging, since it involves both a model for the radial distribution of the first and second populations and how these influence the ultraviolet colours. To give an idea of the potential profiles that may be expected, we assume that the clusters have multiple populations with a similar distribution to those observed in the Galactic globular clusters by \citet{Lardo11}. Specifically, we assume the number of stars in the second population are similar to those in the first population for $r>3r_{h}$ and increase linearly to become three times bigger for $r<1r_{h}$. The resulting change in $FUV-u$ is estimated using the YPES models, where the two populations are assumed to have helium fractions, $Y=0.23$ and 0.33.

The red line in Figure \ref{fig:color_profs} shows this predicted variation in the colour profile. It can be seen that such gradients may be detectable in data with this resolution. However, none of the clusters show the signature of significantly bluer $FUV-u$ colours towards their cores. Surprisingly, B218 is observed to actually have slightly redder colours in its central regions.

We caution that there is significant uncertainty in the plotted prediction and that the profile shown is intended as an example of the possible variation, rather than a comprehensive model. Firstly, the expected fraction and distribution of the populations is not very well constrained by the Galactic clusters and is likely to vary from cluster to cluster. Additionally, the resulting colour change is dependent on both the details of the population abundances and the stellar populations models. We therefore note the lack of any evidence for strong radial variations over the range observed by \citet{Lardo11} in the Galactic globular clusters, but are unable to rule out radially distinct multiple populations. However, we do note that such data are sensitive to radial variations in the HB populations over radial ranges observed in some Galactic globular clusters.

\section{Summary}

We present the ultraviolet surface brightness density profiles of 12 massive globular clusters in M31 based on high resolution \hst photometry. We find no evidence for differences in the surface brightness profiles of the clusters in F140LP-F336W ($FUV-u$). We note that this colour is very sensitive to the temperatures of HB stars in these clusters and that the similar profiles observed suggest no evidence for a population of spatially distinct blue/extreme-HB stars.

Our FUV photometry shows that some of these clusters are brighter in the $FUV$ than expected based on canonical models. This is consistent with similar conclusions based on \galex and \hst observations of other clusters. We consider two scenarios for the production of this extra FUV emission; dynamical interactions enhancing the population of extreme-HB stars and the presence of second populations of helium enhanced HB stars which are hotter than their primordial counterparts.

Dynamically enhanced populations should be formed in the cluster cores, where the stellar density is highest. Given that HB stars have quite short lifetimes, we expect such a population to still be centrally located. The lack of centrally concentrated FUV emission in these clusters therefore argues against a dynamically enhanced extreme-HB population as the origin of the enhanced $FUV$ emission.

The excess FUV emission observed from these clusters is consistent with the clusters hosting second populations of helium enhanced stars. Additionally, we note that the correlation between $FUV-NUV$ and the stellar density/ mass of clusters suggests that the fraction (or helium enhancement) of these second generation stars correlates with these parameters. Such correlations have been observed in the Galactic globular clusters. Our results are therefore consistent with M~31's clusters hosting similar second populations of stars to the Galactic globular clusters. We note that there is no evidence that this second population is centrally concentrated within the clusters.

\section*{Acknowledgements}

We thank the anonymous referee for taking the time to review this paper and for providing prompt and informed feedback. MBP, AK and TJM acknowledge support from HST-GO-14132. MBP and SEZ acknowledge support from the NSF grant AST-1412774. JS acknowledges support from NSF grant AST-1514763 and a Packard Fellowship.

Based on observations made with the NASA/ESA Hubble Space Telescope, obtained from the data archive at the Space Telescope Science Institute. STScI is operated by the Association of Universities for Research in Astronomy, Inc. under NASA contract NAS 5-26555. Support for this work was provided by NASA through grant number HST-GO-14132.001-A from the Space Telescope Science Institute, which is operated by AURA, Inc., under NASA contract NAS 5-26555.




\bibliographystyle{mnras}
\bibliography{bibliography_etal}





\bsp	
\label{lastpage}
\end{document}